# Towards an Adaptive Robot for Sports and Rehabilitation Coaching

Martin K. Ross, Frank Broz, Lynne Baillie

School of Mathematical and Computer Sciences

Heriot-Watt University

Edinburgh

mkr30@hw.ac.uk, f.broz@hw.ac.uk, l.baillie@hw.ac.uk

**Abstract**

The work presented in this paper aims to explore how, and to what extent, an adaptive robotic coach has the potential to provide extra motivation to adhere to long-term rehabilitation and help fill the coaching gap which occurs during repetitive solo practice in high performance sport. Adapting the behavior of a social robot to a specific user, using reinforcement learning (RL), could be a way of increasing adherence to an exercise routine in both domains. The requirements gathering phase is underway and is presented in this paper along with the rationale of using RL in this context.

## Introduction

The field of robotics has evolved since the early days of assembly line manufacturers and is now facing the challenge of interacting with humans in environments which are structured around people rather than machines (e.g. the home). This evolution presents the need to develop technology which sits at the intersection of Artificial Intelligence (AI) and Human-Robot Interaction (HRI). One application area for this type of technology could be a robotic coach for use in both long-term rehabilitation after stroke and individual practice in high performance sport (the exemplar sport we focus on is squash). Both of these activities require the consistent performance of repetitive exercises/drills, often conducted under no supervision over a long period of time. Using a robotic coach could have many benefits including increased engagement and adherence, better feedback, and improved performance of the practiced skills.

Stroke is one of the leading causes of acquired adult disability, affecting 15 million people every year worldwide with more than half of survivors left with permanent disabilities (World Health Organisation 2018) such as fatigue, weakness in the arms and legs, aphasia and forgetfulness (Stroke Association 2018). Adhering to a rehabilitation program and carrying out rehabilitation exercises regularly has the potential to significantly improve a stroke survivor's recovery both in the early days after a stroke and long after they return home (Galvin, Cusack, and Stokes 2009). However, despite the common goal of survivors to improve their overall health and functional ability after stroke, adherence to rehabilitation exercise programs tends to decrease when they are carried out in the home. One of the main reasons for this is survivors' lack of motivation (Jurkiewicz, Marzolini, and Oh 2011).

An area in which motivation to adhere to an exercise routine could be higher is high-performance sport. In sports such as squash independent practice, including repetitive solo drills with little or no input from coaches, is used by many of the top players. In this, we can see parallels between rehabilitation after stroke and training for squash due to the repetitive, unsupervised nature of performing an exercise routine over a long period of time. In other sports, praise for independent practice given by coaches has been shown to increase the intrinsic motivation of athletes (Almagro, Sáenz-López, and Moreno 2010), which can be an important factor in behavioural maintenance (Kilpatrick, Hebert, and Bartholomew 2014). If behaviours such as this praise could be replicated in a Socially Assistive Robot (SAR), it is possible that a robotic coach could have the capability of intrinsically motivating both stroke survivors and athletes to adhere to independent exercise programs.

By creating a robotic coach for this purpose, the following research questions can be addressed:

**R1.** What are the similarities and differences between the strategies used by a robotic coaching system to effectively motivate people in sports and in rehabilitation?

**R2.** How can a social robot be used to increase adherence to exercises in sport and rehabilitation?

**R3.** Does adapting the behaviours of a social robot to a specific user, using reinforcement learning (RL), increase the adherence of that user to an exercise routine?

This paper will provide an overview of the related work in this area and present the work conducted up to this point towards the proposed adaptive robotic coach.

## Background

A stroke is a brain attack which happens when the blood supply to part of the brain is cut off, killing brain cells (Stroke Association 2018). The most common affect is motor impairment: the loss or limitation of functional muscle control or movement, or a limitation in mobility (Langhorne, Coupar, and Pollock 2009). These impairments impact a stroke survivors ability to carry out activities of daily living and engage with technology (Georgiou et al. 2019). The quality of the rehabilitation exercises performed is vital in ensuring that as much improvement as possible is made by the stroke survivor (Levin, Kleim, and Wolf 2009). Therefore, there is a need to design new technology to help this group engage in rehabilitation and provide feedback on the quality of the performed exercise, without the supervision of an expert.

In the current work, squash is considered as a sport in which the independent, repetitive training used by many of the top players has a number of parallels with the nature of motor recovery after stroke. Squash is an intermittent, high-intensity racket sport that is played in a court measuring 9.75 m in length, 6.4 m in width, and 5.64 m in height, with matches contested over the best of 5 games in singles competition (Gibson et al. 2019). There has been research in squash in the area of the physiological demands (Gibson et al. 2019; Murray et al. 2016) and performance requirements (Jones et al. 2018) of the sport, but very little into the coaching techniques used by professionals in this area or the use of technology in the coaching of the sport.

The techniques coaches use to motivate their players is something of particular interest in this project. Extrinsic motivation comes from the environment surrounding a person, whereas intrinsic motivation comes from within. Although extrinsic motivation can have a positive effect on the short-term behaviours of a person towards a specific task, it is intrinsic motivation which is most likely to result in sustained behavioural change over the long term (Fasola and Matarić 2014; Kilpatrick, Hebert, and Bartholomew 2014). Intrinsic motivation can, however, be affected by external factors such as positive information, often in the form of verbal feedback from coaches and teammates (Ryan, Vallerand, and Deci 1985). The greatest sense of intrinsic motivation comes when people are operating at their optimal challenge point (Wade, Parnandi, and Matarić 2011). Therefore, questions arise as to how coaches attempt to affect the intrinsic motivation of their athletes, how they keep them at this optimal challenge point and if it would be possible to provide similar behaviour using a SAR.

The HRI problem is to understand and shape the interactions (the process of working together to accomplish a goal) between one or more humans and one or more robots (Goodrich and Schultz 2007). There can be many parallels seen between the field of HRI and coaching in both sports and rehabilitation. Things such as the communication skills required by social robots (Dautenhahn 2007) are also valuable skills in coaches (Nash, Sproule, and Horton 2011); and the teamwork element of coaching (Knight 2011) is reflected in current research in collaborative robotics (Kragic et al. 2018). These parallels suggest the potential of an HRI system to be used as a rehabilitation or sports coach, and the examples presented in the following sections reinforce this.

It has been suggested that adaptive systems have the potential to motivate users to change their behaviour by personalising their motivational techniques (Kang, Tan, and Miao 2015; Lee et al. 2015). Therefore personalisation of HRI systems will be an important area of interest in this project, and is an area in which RL has potential application. Reinforcement learning is a type of machine learning which learns through interaction with the environment (Sutton and Barto 1998). An agent tries to maximise the reward it gets from its actions on the environment over the long-term life of the system. It uses a policy to decide which actions to undertake given the state of the environment at any given point, and constantly tries to improve the policy through updating estimated rewards for state-action pairs encountered in the interaction. At each timestep, the algorithm controlling the agent must decide whether to exploit its knowledge about the environment and select what appears to be the action giving the highest reward, or explore the environment by selecting an action with a lower predicted reward in an attempt to improve its policy.

## Related Work

### Traditional Technology Use

In rehabilitation robotics, much work has looked at the possibility of exoskeleton devices, or devices which can aid the movement of the impaired limb of a stroke survivor. A review of the literature (Prange et al. 2006) found that up until 2005, the majority of robotic systems designed to aid recovery of the hemiparetic arm after stroke implemented one or more of the following modalities:

- Passive movement – the robot moves the patient's arm.
- Active movement – movement is either partially assisted or resisted by the robot.
- Bimanual exercise – active movement of the unaffected arm mirrored by passive movement of the affected arm.

In comparison, there has been little work in the area of a robotic coach for stroke rehabilitation, which would be capable of, in the absence of a therapist, leading a user through a rehabilitation program, particularly in the home setting.

The use of technology to aid the coaching process has been seen to provide some advantages in sport. In a study in which 12 semi-professional rugby players performed barbell back squats, it was found that providing kinematic feedback either verbally or visually resulted in an almost certain improvement in barbell velocity over a no feedback condition (Weakley et al. 2018). Interestingly, the use of verbal encouragement, with no kinematic feedback, from a coach also increased barbell velocity. Similar effects have been seen when providing augmented feedback (in the form of service speed detected by a speed gun) on improvements in serving speed in high-level junior tennis players (Moran, Murphy, and Marshall 2012). The players were unable to differentiate differences in serve speeds themselves, and the group which received the augmented feedback made significantly bigger improvements over a six week training period than those who did not.

**Coaching with SARs**

Many examples of traditional coaching technology require humans to interpret performance data for themselves. However, there has been some research undertaken on systems which bypass the need for such data analysis from the user, and instead use a robotic coach or companion as a means of interaction. For example, Sussenbach et al. devised a motivation model for a robotic fitness companion (Sussenbach et al. 2014), which was implemented in an autonomous system that guides users through indoor cycle training. It was found that the robotic system was able to engage the user in training over an 18-day study period, and that compliance with instructions relating to the training was higher with the robot, compared to an on-screen condition (Schneider et al. 2015). However, it is not clear in this study if it was actually the presence of the robot itself that contributed to the higher rate of compliance, or if it was down to the fact that the on-screen version only gave instructions, and did not use feedback or motivating behaviours like the robot. Despite its limitations, this study suggests there is potential for this type of SAR to be used in an exercising scenario.

There is some evidence that HRI technology has the potential to deliver this type of coaching in rehabilitation as well, providing many benefits to the user's life. Robotic rehabilitation coaches for children with Cerebral Palsy (CP) have been considered, showing promising levels of engagement and adherence to a rehabilitation program (Malik et al. 2016). The wider population also has the potential to benefit from using a robotic coach to promote physical activity, as demonstrated by Fasola and Matarić (Fasola and Matarić 2014). This study found that an autonomous robotic exercise instructor was successful in motivating elderly users to engage in physical exercise. A robotic coach developed to tackle physical rehabilitation for chronic lower back pain has also been developed in a lab environment (Devanne et al. 2018; Nguyen, Tanguy, and Remy-Neris 2016). Although no full user evaluations of the system have taken place, initial testing indicates that the system is able to identify incorrectly performed exercises and give feedback on the part of the exercise that was wrong.

An autonomous robotic coach, similar to that presented by Fasola and Matarić, has been developed specifically for stroke rehabilitation (Wade, Parnandi, and Matarić 2011). The system uses a humanoid robot torso and head to interact with the stroke survivor and lead them through a rehabilitation session. Although the lab tests carried out on the system show the potential of the use of an autonomous robot to lead a stroke rehabilitation session, the evaluation presented focusses mainly on the efficacy of the wire puzzle task which users were coached through, rather than the effect of the robotic coach itself. Furthermore, there have been no studies identified which evaluate whether a similar setup could be used in the home environment, where rehabilitation has been found to be beneficial (Hillier and Inglis-Jassiem 2010).

**Trust in Human-Robot Interaction**

While working with a robot, particularly in a rehabilitation setting, trust plays a crucial role in establishing and maintaining a patient-agent relationship (Xu, Bryant, and Howard 2018). In a between subjects study which looked into this issue, Xu, Bryant, and Howard found that the level of trust felt by participants towards a robot therapist was comparable to that felt towards a human therapist during the same interaction. Participants in the robot condition actually reported an overall higher level of trust than those in the human therapist condition. However, only 10% of participants had internally questioned the guidance of the robot. Blindly following a robotic system like this could be dangerous when the robot may need to learn a policy through exploration.

A sufficient level of trust would be required if a robotic system was to be deployed in the home, or in the training environment, of a user. The way a robot behaves can have a significant impact on how much a person trusts it. It was found in a study by Correia et al. that a robot displaying group-based emotions in a team-based card game promoted significantly higher levels of group identification and group trust in participants towards the robot compared to one which only expressed individual emotions (Correia et al. 2018). If a similar behavioural model could be implemented in a robotic coach, which saw itself as a teammate of the user, it is possible that sufficient levels of trust would be seen to allow for integration of such a system into the very personal space of someone's home.

**Personalisation using Reinforcement Learning**

To further increase adherence to an exercise program, it may be possible to use RL to adapt the behaviour of a robotic coach to individual users. It has been explored whether RL could be used to personalise the motivational strategies of a robot in teaching scenarios. Roy et al. explored the effect of using participants' performance in a series of pattern memorisation tasks as a way of learning which motivational behaviours work best for that individual (Roy et al. 2018). This RL method was compared to a no-reinforcement (the robot would not give any form of motivation) and a random reinforcement (the robot would randomly choose which motivational behaviour to use) condition, and more effective teaching was observed in the RL condition. Similar results were found by Gordon et al. in the teaching of a second language to children aged 3-5 years old, finding that students who interacted with the robot that personalised its affective feedback strategy showed a significant increase in valence (a measure of the positive or negative nature of the recorded person's experience[1]), as compared to students who interacted with a non-personalising robot (Gordon et al. 2016). The difference between Gordon et al. and Roy et al's work was in the way the reward was determined. The RL algorithm used by Gordon et al's robot utilised the facial expressions of the child to compute the reward function, whereas the participant's performance on the task was used in Roy et al's work.

When using RL in HRI, problems could arise if a limited amount of interaction data is available. Oudah et al investigated this issue, exploring whether two fast-learning online algorithms (counter-factual regret (Johanson et al. 2012) and Gabe-S++ (Crandall 2015)) could learn to collaborate with human partners during 51 rounds of a block-sharing game (Oudah et al. 2015). In this study, the algorithms alone did not consistently learn to collaborate, however humans failed to collaborate with each other without explicit communication as well. When used in conjunction with a cheap-talk (non-binding, unmediated, and costless communication) communication system, the Gabe-S++ algorithm was able to effectively collaborate with people using a limited amount of interaction data.

Speed of convergence is always in the interest of RL algorithms, particularly in real-time HRI scenarios when the user is expecting the robot to provide instant responses. An RL algorithm was combined with Learning from Demonstration to significantly increase the rate of convergence and the speed of task completion (number of steps) in a tea-making activity for people with dementia, when compared to learning cognitive models using Q-learning with an Upper Confidence Bound (UCB) exploration strategy (Moro, Nejat, and Mihailidis 2018). In this system, caregivers could perform a demonstration of behaviours required for a given assistive activity in front of the robot, which would learn to display these behaviours by imitating the combinations of speech and gestures used by the demonstrator. Each behaviour could then be autonomously labelled by the system, which allowed for personalisation of the selection of behaviours to the user's cognitive model (which depended on the new user functioning state, the previous user activity state, and the robot activity state).

But even if algorithms exist which can converge to an optimal policy quickly, HRI scenarios will never have a state set that is fully observable because it is impossible to observe what the human is thinking, and what actions it is planning. So how can we guarantee that the policy learned will actually be a good one? As with Moro, Nejat, and Mihailidis' study, learning from humans could be a way of tackling this issue as well, with Mandel et al. proposing a framework in which a human adds actions to an RL system over time to boost performance (Mandel et al. 2017). Although this framework has not yet been compared to any current state of the art algorithms, and the value of including a human expert in the process remains to be seen, it seems that ideas such as this one would be very useful in HRI scenarios to tackle the data sparsity issue.

Another method which can be used to tackle this problem is to use Bayesian reinforcement learning (BRL). This is an approach to RL that leverages methods from Bayesian inference to incorporate prior information into the learning process, and provides a principled way to tackle the exploration-exploitation problem (Ghavamzadeh et al. 2015). This was a method used by Gordon and Breazeal in an algorithm designed to assess a child's word reading capabilities (Gordon and Breazeal 2015). The algorithm was implemented in a robot and the system tested by 34 children aged 4-8. It was found that the algorithm resulted in an accurate representation of a child's word-reading skills for the whole age range and varying reading skill, indicating that the personalisation of the task by the robot works in allowing child-specific assessment-based tutoring.

Although BRL was only compared to a random condition in this study, it has been shown that using BRL has advantages. These include its ability to tackle the exploration-exploitation problem, to incorporate information into the learning process, and the principled Bayesian approach for andling parameter uncertainty (Ghavamzadeh et al. 2015). However, in Ghavamzadeh et al's review, several limitations are also described: BRL algorithms are often complex and more difficult to implement than their RL counterparts, and are not tractable – solving large problems is still challenging.

Despite these limitations, these studies demonstrate that an approach to personalisation of behaviours of a SAR using RL has the potential to increase both the performance and enjoyment of users of the system, thus supporting the use of such a method in the current work.

---

[1] Definition given by Affectiva (https://developer.affectiva.com/metrics/) – developers of the software used in Gordon et al's study for automatically analysing facial expressions in real time.

## Methodology

The design of the robotic coaching system proposed in this paper is not trivial, so the work conducted to date has focused on requirements gathering. It has taken the form of a Participatory Design (PD) workshop and a systematic observation study of the behaviours of squash coaches (underway at time of writing).

### Participatory Design Workshop

No studies could be identified considering Participatory Design (PD) with stroke survivors and home-based, self-managed rehabilitation using SARs. Given this, and the success of PD techniques in the field of Human Computer Interaction (HCI), it was decided to run a design workshop to address this issue. Details of the workshop have previously been published (Georgiou et al. 2019) so only an overview of the results will be given here for context.

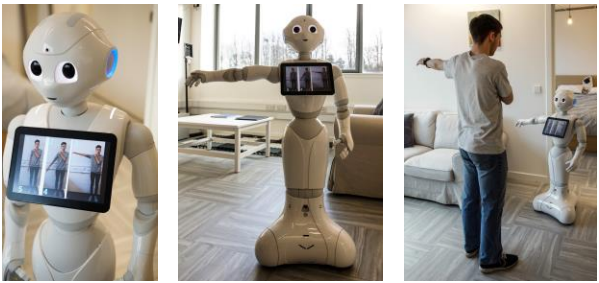

Figure 1 - Pepper demonstrating an upper limb rehabilitation exercise. (Georgiou et al. 2019)

It was found that the Pepper robot (Figure 1), whilst speaking at quite a slow and even pace, spoke much too quickly for a stroke survivor. Many assistive robot designs (and assistive conversational agents like Alexa), use speech as the primary interface, thus existing assistive agent designs often cannot just be re-used for stroke survivors. They may need non-speech interfaces (or the speech interface may need a very different design).

A wide variety of human-robot interactions were designed by participants, from purely utilitarian robots to those with agency. Of particular interest to this work are the designs in Figure 2 and Figure 3 which represent rehabilitation coaches. Figure 3 shows Pepper being used as a demonstrator of rehabilitation exercises, whereas Figure 2 indicates that a display of the exercise could appear on Pepper's tablet computer screen, but could also be connected to a TV or computer monitor. Both designs also indicated that the system should give reminders of when to exercise, but also persist if the reminder was "delayed" until the exercises were completed. The robot would be able to tell that the user was doing the exercise, and feedback this information to physiotherapists. All of these suggestions are things which will be taken on board in the design of the robotic coach to be used in this project.

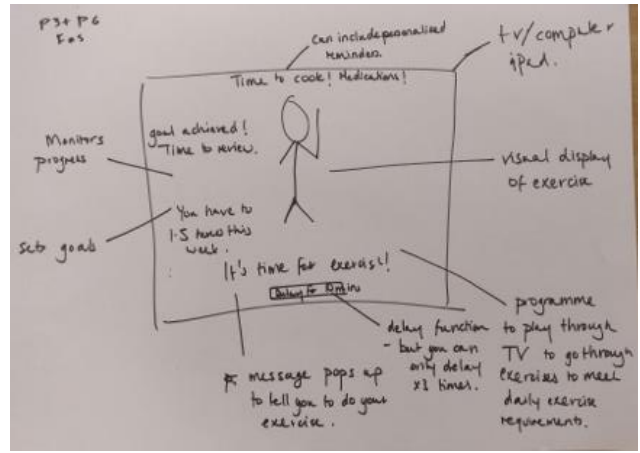

Figure 2 - Design for Pepper's chest tablet screen. (Georgiou et al. 2019)

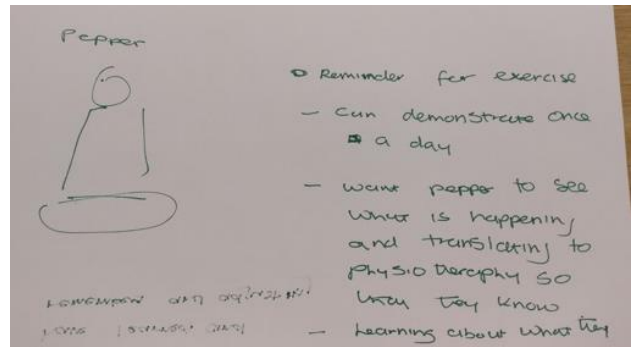

Figure 3 - Design of a robot as a rehabilitation coach. (Georgiou et al. 2019)

### Systematic Observations of Squash Coaches' Behaviour

In order to help answer **R1**, and to enable the design of a robotic system which can carry out the role of a rehabilitation or squash coach, it must first be understood how professionals in these two domains behave so that realistic behaviours can be implemented in the robot. The sports coaching research community sees systematic observation as a valuable tool in developing a greater understanding of what coaches do in practice and competition (Cope, Partington, and Harvey 2017). However, no studies have been identified which use a systematic observation instrument to study the behaviours of professional squash coaches. In fact, in a recent review of the literature surrounding systematic observation in sport, Cope, Partington, and Harvey identified only one study between 1997 and 2016 which observed the behaviours of coaches in an individual sport (golf) (Cope, Partington, and Harvey 2017). Therefore this study, which is currently underway and consists of systematic observations of 6-10 professional squash coaches conducting 2 individual coaching sessions each, has the potential to fill this gap in the literature.

Due to the study not yet being completed (5 sessions have been undertaken at time of writing), a full report will not be given here. However, it is intended that the following aims will be addressed:

- **A1.** Gather and analyse data about the behaviours of squash coaches to discover the most prevalent behaviours during coaching sessions and during which segments of the coaching session they occur.
  - **A1.1.** Use this data to formalise semi-structured interview questions which will be used to understand why professionals in this area behave in the ways that they do.
  - **A1.2.** Use this data as comparison between the behaviours of squash coaches and rehabilitation therapists.
  - **A1.3.** Use this data along with the interview answers and observations/interviews of rehabilitation therapists to design features of a robotic squash/rehabilitation coach.
- **A2.** Discover any differences in behaviours of coaches based on factors such as experience of player and length of time working with player.

The observation instrument being used in this study is a modified version of the ASUOI (Lacy and Darst 1984). The aim when designing this instrument was to keep it as simple as possible, while collecting enough information to satisfy the objectives outlined above. The instrument allows the observed behaviours of a coach to be coded into categories for later analysis. Categories include pre-, concurrent and post-instruction, positive and negative modelling, and praise.

The coding is being performed live during the session for the majority of participants. However, to obtain inter-observer reliability and perform the recommended observation coder training (Cope, Partington, and Harvey 2017), 5 of the sessions have been video and audio recorded. This has been done using an iPhone 5s' 8-megapixel camera connected via Bluetooth to a Bb Talkin' Advance microphone attached to the coach's clothing.

Due to a new observation instrument being used for this study, a similar process is being followed to that used in (Hall, Gray, and Sproule 2016). Firstly, the lead observer familiarised himself with the ASUOI through careful reading of the literature. Next, face validity of the developed instrument was obtained through review from a top-level squash coach (currently a National coach) and an experienced coder. Repeated practice with the developed instrument, using both video footage and live sessions of professional squash coaches is currently being undertaken, with gaps of 24 hours, 7 days and 14 days planned to allow for memory lapse (Lacy and Darst 1989). Coder training from an experienced coder has also taken place.

Reliability of the observations will be achieved via a consensus-based approach. The 5 recorded sessions have been coded by the lead observer, and the coding will be validated by an experienced observer. Discussions will then take place between the two observers on areas where there is any disagreement so that minimal individual interpretation of behaviours can take place, and subsequent sessions are coded in agreement with both observers. This will provide sufficient inter-observer reliability for the lead observer to conduct the remainder of the sessions alone.

Once observations are complete, the data will be analysed in order to address the aims of the study. This will allow the discovery of the actions that make up coaching behaviour in this context. By implementing a version of these actions in a SAR, an RL agent will then be able to personalise the robot's behavioural strategy by selecting the best action to perform at a given time based on the user it is trying to motivate.

## Overview of Proposed System

The requirements of the system to be developed in the current work are yet to be finalised due to the ongoing nature of the above observation study and the planned studies mentioned in the Future Work section. However, a speculative overview of the planned functionalities of the system are presented here.

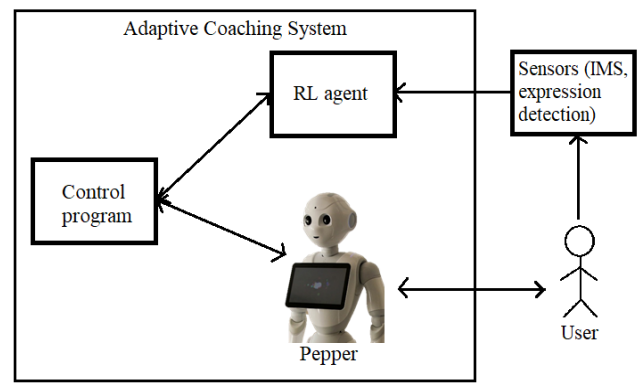

Figure 4 - Architecture diagram of proposed system.

The system will be developed for use in both the home environment of a stroke survivor and the training environment of a high level squash player (i.e. squash court), using the Pepper robot as the hardware. A high-level architecture diagram is shown in Figure 4.

The RL agent will use information obtained from sensors (most likely an inertial motion sensor (IMS) to detect movement and a facial expression recogniser as used in (Gordon et al. 2016)) and use this to update its policy on the best behaviour to select at each timestep. This decision will be used by the control program to animate Pepper in an appropriate way. Pepper will act as the coach – becoming an interface between the control and the user.

# Future Work

Following completion of the systematic observation study of professional squash coaches' behaviours, similar observations, using the same observation instrument, will be conducted on physiotherapists leading stroke survivors through rehabilitation sessions. The data obtained from these observation studies will be analysed and used to compare the behaviours of professionals in both fields and synthesize questions to be used in semi-structured interviews with the observed squash coaches and rehabilitation therapists. The interview study will provide a deeper understanding of why the participants use the behaviours they do, and help the observer clarify anything which may have come up during the observations.

Using the analysed data from these studies, the requirements for an autonomous squash and stroke rehabilitation coach will be formalised and the system will be implemented. There will be challenges in both the HRI and RL aspects of implementation but it is hoped that using an iterative design and implementation process with heavy involvement from end users throughout the process, as used in the development of other systems designed to aid stroke survivors (Alankus et al. 2010; Ozgur et al. 2018), will help tackle the interaction side of this. It is yet to be explored exactly how the RL agent will be implemented, but Gordon et al. and Roy et al's work will provide a good starting point from which to build on (Gordon et al. 2016; Roy et al. 2018).

The end goal is to conduct a long-term deployment of the system in the target environment to evaluate all three research questions posed in the Introduction.

# Acknowledgments

We thank all participants for their collaboration in the design workshop and the observation sessions. We also thank Chest Heart & Stroke Scotland and Scottish Squash for all their help during the participant recruitment phase and during both studies. This work was supported by Engineering and Physical Sciences Research Council (EPSRC), Grant.ID: EPSRC DTP18.